\documentclass[twocolumn,pre,superscriptaddress]{revtex4}

\usepackage{dcolumn}
\usepackage{amsmath}

\usepackage{graphicx}

\begin{document}

\renewcommand{\d}{d}
\newcommand{\Ord}{\mathrm{O}}
\newcommand{\eref}[1]{(\ref{#1})}
\newcommand{\etal}{{\it{}et~al.}}
\newcommand{\defn}{\textit}
\newcommand{\half}{\mbox{$\frac12$}}

\newlength{\figurewidth}
\setlength{\figurewidth}{0.95\columnwidth}
\setlength{\parskip}{0pt}
\setlength{\tabcolsep}{6pt}
\setlength{\arraycolsep}{2pt}

\title{Finding community structure in very large networks}
\author{Aaron Clauset}
\affiliation{Department of Computer Science, University of New Mexico,
Albuquerque, NM 87131}
\author{M. E. J. Newman}
\affiliation{Department of Physics and Center for the Study of Complex
Systems,\\
University of Michigan, Ann Arbor, MI 48109}
\author{Cristopher Moore}
\affiliation{Department of Computer Science, University of New Mexico,
Albuquerque, NM 87131}
\affiliation{Department of Physics and Astronomy, University of New Mexico,
Albuquerque, NM 87131}

\begin{abstract}
The discovery and analysis of community structure in networks is a topic of
considerable recent interest within the physics community, but most methods
proposed so far are unsuitable for very large networks because of their
computational cost.  Here we present a hierarchical agglomeration algorithm
for detecting community structure which is faster than many competing
algorithms: its running time on a network with $n$ vertices and $m$ edges
is $\Ord(m d \log n)$ where $d$ is the depth of the dendrogram describing
the community structure.  Many real-world networks are sparse and
hierarchical, with $m \sim n$ and $d \sim \log n$, in which case our
algorithm runs in essentially linear time, $\Ord(n \log^2 n)$.  As an
example of the application of this algorithm we use it to analyze a network
of items for sale on the web-site of a large online retailer, items in the
network being linked if they are frequently purchased by the same buyer.
The network has more than $400\,000$ vertices and 2 million edges.  We show
that our algorithm can extract meaningful communities from this network,
revealing large-scale patterns present in the purchasing habits of
customers.
\end{abstract}
\maketitle

\section{Introduction}
Many systems of current interest to the scientific community can usefully
be represented as networks~\cite{Strogatz01,AB02,DM02,Newman03d}.  Examples
include the Internet~\cite{FFF99} and the world-wide
web~\cite{AJB99,Kleinberg99b}, social networks~\cite{WF94}, citation
networks~\cite{Price65,Redner98}, food webs~\cite{DWM02a}, and biochemical
networks~\cite{Kauffman69,Ito01}.  Each of these networks consists of a set
of nodes or \defn{vertices} representing, for instance, computers or
routers on the Internet or people in a social network, connected together
by links or \defn{edges}, representing data connections between computers,
friendships between people, and so forth.

One network feature that has been emphasized in recent work is
\defn{community structure}, the gathering of vertices into groups such that
there is a higher density of edges within groups than between
them~\cite{note}.  The problem of detecting such communities within
networks has been well studied.  Early approaches such as the
Kernighan--Lin algorithm~\cite{KL70}, spectral
partitioning~\cite{Fiedler73,PSL90}, or hierarchical
clustering~\cite{Scott00} work well for specific types of problems
(particularly graph bisection or problems with well defined vertex
similarity measures), but perform poorly in more general
cases~\cite{Newman04b}.

To combat this problem a number of new algorithms have been proposed in
recent years.  Girvan and Newman~\cite{GN02,NG04} proposed a divisive
algorithm that uses edge betweenness as a metric to identify the boundaries
of communities.  This algorithm has been applied successfully to a variety
of networks, including networks of email messages, human and animal social
networks, networks of collaborations between scientists and musicians,
metabolic networks and gene
networks~\cite{GN02,GN04,Guimera03,HHJ03,HH03,TWH03,GD03,BPDA04,WH04b,Arenas04}.
However, as noted in~\cite{NG04}, the algorithm makes heavy demands on
computational resources, running in $\Ord(m^2 n)$ time on an arbitrary
network with $m$ edges and $n$ vertices, or $\Ord(n^3)$ time on a sparse
graph (one in which $m\sim n$, which covers most real-world networks of
interest).  This restricts the algorithm's use to networks of at most a few
thousand vertices with current hardware.

More recently a number of faster algorithms have been
proposed~\cite{Radicchi04,Newman04a,WH04a}.  In~\cite{Newman04a}, one of us
proposed an algorithm based on the greedy optimization of the quantity
known as \defn{modularity}~\cite{NG04}.  This method appears to work well
both in contrived test cases and in real-world situations, and is
substantially faster than the algorithm of Girvan and Newman.  A naive
implementation runs in time $\Ord((m+n)n)$, or $\Ord(n^{2})$ on a sparse
graph.

Here we propose a new algorithm that performs the same greedy optimization
as the algorithm of~\cite{Newman04a} and therefore gives identical results
for the communities found.  However, by exploiting some shortcuts in the
optimization problem and using more sophisticated data structures, it runs
far more quickly, in time $\Ord(m d \log n)$ where $d$ is the depth of the
``dendrogram'' describing the network's community structure.  Many
real-world networks are sparse, so that $m \sim n$; and moreover, for
networks that have a hierarchical structure with communities at many
scales, $d \sim \log n$.  For such networks our algorithm has essentially
linear running time, $\Ord(n \log^2 n)$.

This is not merely a technical advance but has substantial practical
implications, bringing within reach the analysis of extremely large
networks.  Networks of ten million vertices or more should be possible in
reasonable run times.  As an example, we give results from the application
of the algorithm to a recommender network of books from the online
bookseller Amazon.com, which has more than $400\,000$ vertices and two
million edges.

\section{The algorithm}
\defn{Modularity}~\cite{NG04} is a property of a network and a specific
proposed division of that network into communities.  It measures when the
division is a good one, in the sense that there are many edges within
communities and only a few between them.  Let $A_{vw}$ be an element of the
adjacency matrix of the network thus:
\begin{equation}
A_{vw} = \biggl\lbrace\begin{array}{ll}
           1 & \quad\mbox{if vertices $v$ and $w$ are connected,}\\
           0 & \quad\mbox{otherwise.}
         \end{array}
\end{equation}
and suppose the vertices are divided into communities such that vertex~$v$
belongs to community~$c_v$.  Then the fraction of edges that fall within
communities, i.e.,~that connect vertices that both lie in the same
community, is
\begin{equation}
{\sum_{vw} A_{vw} \delta(c_v,c_w)\over\sum_{vw} A_{vw}}
  = {1\over2m} \sum_{vw} A_{vw} \delta(c_v,c_w),
\end{equation}
where the $\delta$-function $\delta(i,j)$ is 1 if $i=j$ and 0 otherwise,
and $m=\half\sum_{vw} A_{vw}$ is the number of edges in the graph.  This
quantity will be large for good divisions of the network, in the sense of
having many within-community edges, but it is not, on its own, a good
measure of community structure since it takes its largest value of~1 in the
trivial case where all vertices belong to a single community.  However, if
we subtract from it the expected value of the same quantity in the case of
a randomized network, we do get a useful measure.

The \defn{degree}~$k_v$ of a vertex~$v$ is defined to be the number of
edges incident upon it:
\begin{equation}
k_v = \sum_w A_{vw}.
\end{equation}
The probability of an edge existing between vertices $v$ and $w$ if
connections are made at random but respecting vertex degrees is $k_v
k_w/2m$.  We define the modularity~$Q$ to be
\begin{equation}
Q = {1\over2m} \sum_{vw} \biggl[ A_{vw} - {k_v k_w\over2m} \biggr]
    \delta(c_v,c_w).
\label{modularity}
\end{equation}
If the fraction of within-community edges is no different from what we
would expect for the randomized network, then this quantity will be zero.
Nonzero values represent deviations from randomness, and in practice it is
found that a value above about 0.3 is a good indicator of significant
community structure in a network.

If high values of the modularity correspond to good divisions of a network
into communities, then one should be able to find such good divisions by
searching through the possible candidates for ones with high modularity.
While finding the global maximum modularity over all possible divisions
seems hard in general, reasonably good solutions can be found with
approximate optimization techniques.  The algorithm proposed
in~\cite{Newman04a} uses a greedy optimization in which, starting with each
vertex being the sole member of a community of one, we repeatedly join
together the two communities whose amalgamation produces the largest
increase in~$Q$.  For a network of $n$ vertices, after $n-1$ such joins we
are left with a single community and the algorithm stops.  The entire
process can be represented as a tree whose leaves are the vertices of the
original network and whose internal nodes correspond to the joins.  This
\defn{dendrogram} represents a hierarchical decomposition of the network
into communities at all levels.

The most straightforward implementation of this idea (and the only one
considered in~\cite{Newman04a}) involves storing the adjacency matrix of
the graph as an array of integers and repeatedly merging pairs of rows and
columns as the corresponding communities are merged.  For the case of the
sparse graphs that are of primary interest in the field, however, this
approach wastes a good deal of time and memory space on the storage and
merging of matrix elements with value~0, which is the vast majority of the
adjacency matrix.  The algorithm proposed in this paper achieves speed (and
memory efficiency) by eliminating these needless operations.

To simplify the description of our algorithm let us define the following
two quantities:
\begin{equation}
e_{ij} = {1\over2m} \sum_{vw} A_{vw} \delta(c_v,i) \delta(c_w,j),
\end{equation}
which is the fraction of edges that join vertices in community~$i$ to
vertices in community~$j$, and
\begin{equation}
a_i = {1\over2m} \sum_v k_v\delta(c_v,i),
\end{equation}
which is the fraction of ends of edges that are attached to vertices in
community~$i$.  Then, writing
$\delta(c_v,c_w)=\sum_i \delta(c_v,i) \delta(c_w,i)$, we have, from
Eq.~\eref{modularity}
\begin{eqnarray}
Q &=& {1\over2m} \sum_{vw} \biggl[ A_{vw} - {k_v k_w\over2m} \biggr]
      \sum_i\delta(c_v,i)\delta(c_w,i)\nonumber\\
  &=& \sum_i \biggl[ {1\over2m} \sum_{vw} A_{vw} \,\delta(c_v,i)\delta(c_w,i)
        \nonumber\\
  & & \qquad {} - {1\over2m}\sum_v k_v \,\delta(c_v,i)
           {1\over2m}\sum_w k_w\delta(c_w,i) \biggr]\nonumber\\
  &=& \sum_i (e_{ii} - a_i^2).
\end{eqnarray}

The operation of the algorithm involves finding the changes in $Q$ that
would result from the amalgamation of each pair of communities, choosing
the largest of them, and performing the corresponding amalgamation.  One
way to envisage (and implement) this process is to think of network as a
multigraph, in which a whole community is represented by a vertex, bundles
of edges connect one vertex to another, and edges internal to communities
are represented by self-edges.  The adjacency matrix of this multigraph has
elements$A'_{ij} = 2m e_{ij}$, and the joining of two communities $i$ and
$j$ corresponds to replacing the $i$th and $j$th rows and columns by their
sum.  In the algorithm of~\cite{Newman04a} this operation is done
explicitly on the entire matrix, but if the adjacency matrix is sparse
(which we expect in the early stages of the process) the operation can be
carried out more efficiently using data structures for sparse matrices.
Unfortunately, calculating $\Delta Q_{ij}$ and finding the pair $i,j$ with
the largest $\Delta Q_{ij}$ then becomes time-consuming.

In our new algorithm, rather than maintaining the adjacency matrix and
calculating~$\Delta Q_{ij}$, we instead maintain and update a matrix of
value of $\Delta Q_{ij}$.  Since joining two communities with no edge
between them can never produce an increase in~$Q$, we need only store
$\Delta Q_{ij}$ for those pairs $i,j$ that are joined by one or more edges.
Since this matrix has the same support as the adjacency matrix, it will be
similarly sparse, so we can again represent it with efficient data
structures.  In addition, we make use of an efficient data structure to
keep track of the largest $\Delta Q_{ij}$.  These improvements result in a
considerable saving of both memory and time.

In total, we maintain three data structures:
\begin{enumerate}
\item A sparse matrix containing $\Delta Q_{ij}$ for each pair $i,j$ of
communities with at least one edge between them.  We store each row of the
matrix both as a balanced binary tree (so that elements can be found or
inserted in $O(\log n)$ time) and as a max-heap (so that the largest
element can be found in constant time).
\item A max-heap $H$ containing the largest element of each row of the
matrix $\Delta Q_{ij}$ along with the labels $i,j$ of the corresponding
pair of communities.
\item An ordinary vector array with elements~$a_i$.
\end{enumerate}

As described above we start off with each vertex being the sole member of a
community of one, in which case $e_{ij}=1/2m$ if $i$ and $j$ are connected
and zero otherwise, and $a_i=k_i/2m$.  Thus we initially set
\begin{equation}
\label{eq:qinit}
\Delta Q_{ij} = \biggl\lbrace\begin{array}{ll}
                  1/2m - k_i k_j/(2m)^2 &
                    \quad\mbox{if $i,j$ are connected,}\\
                  0 & \quad\mbox{otherwise,}
                \end{array}
\end{equation}
and
\begin{equation}
\label{eq:ainit} 
a_i = \frac{k_i}{2m} 
\end{equation}
for each~$i$.  (This assumes the graph is unweighted; weighted graphs are a
simple generalization~\cite{Newman05a}.)

Our algorithm can now be defined as follows.
\begin{enumerate}
\item Calculate the initial values of $\Delta Q_{ij}$ and $a_i$ according
to~\eref{eq:qinit} and~\eref{eq:ainit}, and populate the max-heap with the
largest element of each row of the matrix $\Delta Q$.
\item Select the largest $\Delta Q_{ij}$ from $H$, join the corresponding
communities, update the matrix $\Delta Q$, the heap $H$ and $a_i$ (as
described below) and increment $Q$ by $\Delta Q_{ij}$.
\item Repeat step 2 until only one community remains.
\end{enumerate}

Our data structures allow us to carry out the updates in step 2 quickly.
First, note that we need only adjust a few of the elements of $\Delta Q$.
If we join communities $i$ and~$j$, labeling the combined community~$j$,
say, we need only update the $j$th row and column, and remove the $i$th row
and column altogether.  The update rules are as follows.
\begin{subequations}
If community $k$ is connected to both $i$ and $j$, then
\begin{equation}
\label{eq:both}
\Delta Q'_{jk} = \Delta Q_{ik} + \Delta Q_{jk} 
\end{equation}
If $k$ is connected to $i$ but not to~$j$, then
\begin{equation}
\label{eq:justi}
\Delta Q'_{jk} = \Delta Q_{ik} - 2 a_j a_k 
\end{equation}
If $k$ is connected to $j$ but not to $i$, then
\begin{equation}
\label{eq:justj}
\Delta Q'_{jk} = \Delta Q_{jk} - 2 a_i a_k.
\end{equation}
\end{subequations}
Note that these equations imply that $Q$ has a single peak over the course of the algorithm, since after the largest $\Delta Q$ becomes negative all the $\Delta Q$ can only decrease.

To analyze how long the algorithm takes using our data structures, let us
denote the degrees of $i$ and $j$ in the reduced graph---i.e.,~the numbers
of neighboring communities---as $|i|$ and $|j|$ respectively.  The first
operation in a step of the algorithm is to update the $j$th row.  To
implement Eq.~\eref{eq:both}, we insert the elements of the $i$th row into
the $j$th row, summing them wherever an element exists in both columns.
Since we store the rows as balanced binary trees, each of these $|i|$
insertions takes $O(\log |j|) \le O(\log n)$ time.  We then update the
other elements of the $j$th row, of which there are at most $|i|+|j|$,
according to Eqs.~\eref{eq:justi} and~\eref{eq:justj}.  In the $k$th row,
we update a single element, taking $O(\log |k|) \le O(\log n)$ time, and
there are at most $|i|+|j|$ values of $k$ for which we have to do this.
All of this thus takes $O((|i|+|j|) \log n)$ time.

We also have to update the max-heaps for each row and the overall max-heap
$H$.  Reforming the max-heap corresponding to the $j$th row can be done in
$O(|j|)$ time~\cite{CLRS01}.  Updating the max-heap for the $k$th row by
inserting, raising, or lowering $\Delta Q_{kj}$ takes $O(\log |k|) \le
O(\log n)$ time.  Since we have changed the maximum element on at most
$|i|+|j|$ rows, we need to do at most $|i|+|j|$ updates of $H$, each of
which takes $O(\log n)$ time, for a total of $O((|i|+|j|) \log n)$.

Finally, the update $a'_j = a_j + a_i$ (and $a_i = 0$) is trivial and can
be done in constant time.

Since each join takes $O((|i|+|j|) \log n)$ time, the total running time is
at most $O(\log n)$ times the sum over all nodes of the dendrogram of the
degrees of the corresponding communities.  Let us make the worst-case
assumption that the degree of a community is the sum of the degrees of all
the vertices in the original network comprising it.  In that case, each
vertex of the original network contributes its degree to all of the
communities it is a part of, along the path in the dendrogram from it to
the root.  If the dendrogram has depth~$d$, there are at most $d$ nodes in
this path, and since the total degree of all the vertices is~$2m$, we have
a running time of $O(m d \log n)$ as stated.

We note that, if the dendrogram is unbalanced, some time savings can be
gained by inserting the sparser row into the less sparse one.  In addition,
we have found that in practical situations it is usually unnecessary to
maintain the separate max-heaps for each row.  These heaps are used to find
the largest element in a row quickly, but their maintenance takes a
moderate amount of effort and this effort is wasted if the largest element
in a row does not change when two rows are amalgamated, which turns out
often to be the case.  Thus we find that the following simpler
implementation works quite well in realistic situations: if the largest
element of the $k$th row was $\Delta Q_{ki}$ or $\Delta Q_{kj}$ and is now
reduced by Eq.~\eref{eq:justi} or~\eref{eq:justj}, we simply scan the $k$th
row to find the new largest element. Although the worst-case running time
of this approach has an additional factor of~$n$, the average-case running
time is often better than that of the more sophisticated algorithm. It should be noted 
that the hierarchies generated by these two versions of our algorithm will 
differ slightly as a result of the differences in how ties are broken for the 
maximum element in a row. However, we find that in practice these differences 
do not cause significant deviations in the modularity, the community 
size distribution, or the composition of the largest communities.

\begin{figure}[t]
\begin{center}
\includegraphics[scale=0.45]{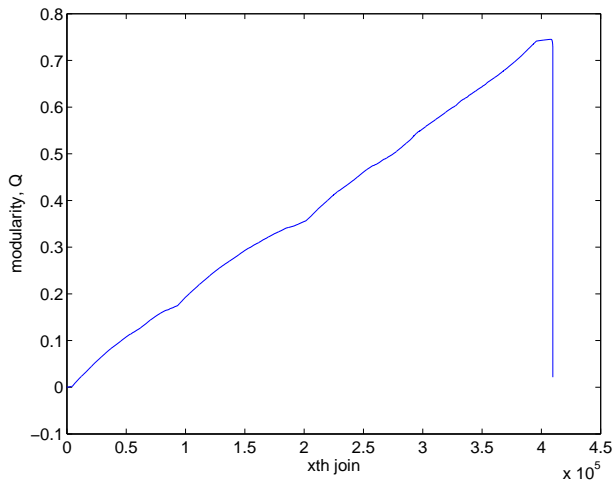}
\end{center}
\caption{The modularity $Q$ over the course of the algorithm (the $x$ axis
shows the number of joins). Its maximum value is $Q=0.745$, where the
partition consists of $1684$ communities.}
\label{fig:Q}
\end{figure}

\section{Amazon.com purchasing network}
The output of the algorithm described above is precisely the same as that
of the slower hierarchical algorithm of~\cite{Newman04a}.  The much
improved speed of our algorithm however makes possible studies of very
large networks for which previous methods were too slow to produce useful
results.  Here we give one example, the analysis of a co-purchasing or
``recommender'' network from the online vendor Amazon.com.  Amazon sells a
variety of products, particularly books and music, and as part of their web
sales operation they list for each item~A the ten other items most
frequently purchased by buyers of~A.  This information can be represented
as a directed network in which vertices represent items and there is a edge
from item~A to another item~B if B was frequently purchased by buyers of~A.
In our study we have ignored the directed nature of the network (as is
common in community structure calculations), assuming any link between two
items, regardless of direction, to be an indication of their similarity.
The network we study consists of items listed on the Amazon web site in
August 2003.  We concentrate on the largest component of the network, which
has $409\,687$ items and $2\,464\,630$ edges.

\begin{figure}[t]
\begin{center}
\includegraphics[width=3in]{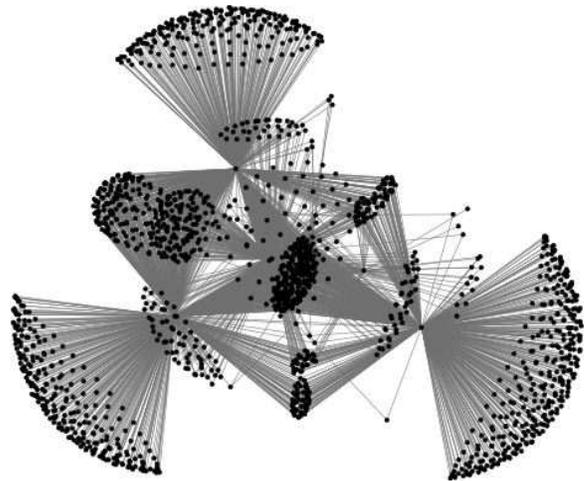}
\end{center}
\caption{A visualization of the community structure at maximum modularity.
Note that the some major communities have a large number of ``satellite''
communities connected only to them (top, lower left, lower right).  Also,
some pairs of major communities have sets of smaller communities that act
as ``bridges'' between them (e.g., between the lower left and lower right,
near the center).}
\label{fig:visual}
\end{figure}

The dendrogram for this calculation is of course too big to draw, but
Fig.~\ref{fig:Q} illustrates the modularity over the course of the
algorithm as vertices are joined into larger and larger groups.  The
maximum value is $Q=0.745$, which is high as calculations of this type
go~\cite{NG04,Newman04a} and indicates strong community structure in the
network.  The maximum occurs when there are $1684$ communities with a mean
size of $243$ items each.  Fig.~\ref{fig:visual} gives a visualization of
the community structure, including the major communities, smaller
``satellite'' communities connected to them, and ``bridge'' communities
that connect two major communities with each other.

\begin{table*}
\begin{center}
\begin{tabular}{cr|p{14.5cm}}
Rank & Size & Description \\
\hline
1 & 114538 & General interest: politics; art/literature; general fiction;
human nature; technical books; how things, people, computers, societies
work, etc. \\

2 & 92276 & The arts: videos, books, DVDs about the creative and performing
arts \\

3 & 78661 & Hobbies and interests I: self-help; self-education; popular
science fiction, popular fantasy; leisure; etc. \\

4 & 54582 & Hobbies and interests II: adventure books; video games/comics;
some sports; some humor; some classic fiction; some western religious
material; etc. \\

5 & 9872 & classical music and related items \\

6 & 1904 & children's videos, movies, music and books \\

7 & 1493 & church/religious music; African-descent cultural books;
homoerotic imagery \\

8 & 1101 & pop horror; mystery/adventure fiction \\

9 & 1083 & jazz; orchestral music; easy listening \\

10 & 947 & engineering; practical fashion 
\end{tabular}
\end{center}\caption{The $10$ largest communities in the
Amazon.com network, which account for $87\%$ of the vertices in the
network.}
\label{table:labels}
\end{table*}

Looking at the largest communities in the network, we find that they tend
to consist of items (books, music) in similar genres or on similar topics.
In Table~\ref{table:labels}, we give informal descriptions of the ten
largest communities, which account for about 87\% of the entire network. 
The remainder is generally divided into small, densely connected 
communities that represent highly specific co-purchasing habits, e.g.,~major 
works of science fiction ($162$ items), music by John Cougar Mellencamp 
($17$ items), and books about (mostly female) spies in the American Civil 
War ($13$ items).  It is worth noting that because few real-world 
networks have community metadata associated with them to which we 
may compare the inferred communities, this type of manual check of
the veracity and coherence of the algorithm's output is often necessary.

\begin{figure}[t]
\begin{center}
\includegraphics[scale=0.45]{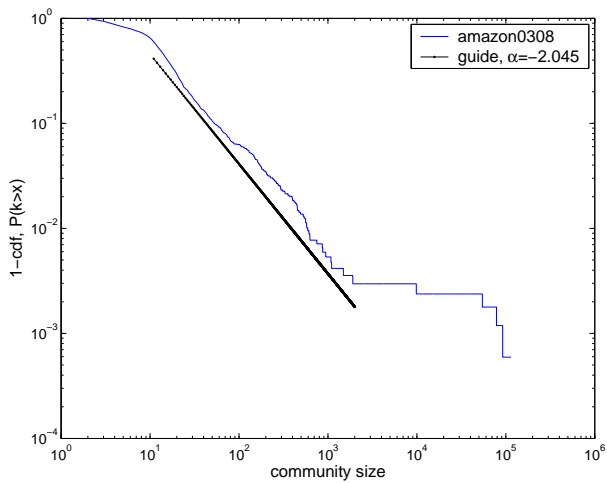}
\end{center}
\caption{Cumulative distribution of the sizes of communities when the
network is partitioned at the maximum modularity found by the algorithm.
The distribution appears to follow a power law form over two decades in the
central part of its range, although it deviates in the tail.  As a guide to
the eye, the straight line has slope $-1$, which corresponds to an exponent
of $\alpha=2$ for the raw probability distribution.}
\label{fig:distribution}
\end{figure}

One interesting property recently noted in some
networks~\cite{Arenas04,Newman04a} is that when partitioned at the point of
maximum modularity, the distribution of community sizes~$s$ appears to have
a power-law form $P(s) \sim s^{-\alpha}$ for some constant~$\alpha$, at
least over some significant range.  The Amazon co-purchasing network also
seems to exhibit this property, as we show in Fig.~\ref{fig:distribution},
with an exponent $\alpha\simeq2$.  It is unclear why such a distribution
should arise, but we speculate that it could be a result either of the
sociology of the network (a power-law distribution in the number of people
interested in various topics) or of the dynamics of the community structure
algorithm.  We propose this as a direction for further research.

\section{Conclusions}
We have described a new algorithm for inferring community structure from
network topology which works by greedily optimizing the modularity.  Our
algorithm runs in time $\Ord(m d \log n)$ for a network with $n$ vertices
and $m$ edges where $d$ is the depth of the dendrogram.  For networks that
are hierarchical, in the sense that there are communities at many scales
and the dendrogram is roughly balanced, we have $d \sim \log n$.  If the
network is also sparse, $m \sim n$, then the running time is essentially
linear, $\Ord(n \log^2 n)$.  This is considerably faster than most previous
general algorithms, and allows us to extend community structure analysis to
networks that had been considered too large to be tractable.  
We have demonstrated our algorithm with an application to a large network 
of co-purchasing data from the online retailer Amazon.com.  
Our algorithm discovers clear communities within this network
that correspond to specific topics or genres of books or music, indicating
that the co-purchasing tendencies of Amazon customers are strongly
correlated with subject matter.  Our algorithm should allow researchers to
analyze even larger networks with millions of vertices and tens of millions
of edges using current computing resources, and we look forward to seeing
such applications.


\begin{acknowledgements}
The authors are grateful to Amazon.com and Eric Promislow for providing the
purchasing network data.  This work was funded in part by the National
Science Foundation under grant PHY-0200909 (AC, CM) and by a grant from
the James S. McDonell Foundation (MEJN).
\end{acknowledgements}


\begin{thebibliography}{10}
\bibitem{note}
Community structure is sometimes referred to as ``clustering'' in sociology
or computer science, but this term is commonly used to mean something else
in the physics literature~\cite{WS98}, so to prevent confusion we avoid it
here.  We note also that the problem of finding communities in a network is
somewhat ill-posed, since we haven't defined precisely what a community is.
A number of definitions have been proposed~\cite{WF94,FLGC02,Radicchi04},
but none is standard.
\end{thebibliography}

\end{document}